\documentclass[12pt]{article}
\usepackage{amsmath}
\usepackage{graphicx}
\begin{document}

\title{\textbf{{Symmetric correlations as seen at RHIC} }}
\author{Adam Bzdak\\
RIKEN BNL Research Center\\
Brookhaven National Laboratory\thanks{
Address: Upton, NY 11973, USA; email: abzdak@bnl.gov}}
\date{}
\maketitle

\begin{abstract}
We analyze the forward-backward multiplicity correlation coefficient as
measured by STAR. We show that in the most central Au+Au collisions bins
located symmetrically around $\eta =0$ with large separation in
pseudorapidity are more strongly correlated than bins located asymmetrically
with smaller separation. In proton-proton collisions the opposite effect is
observed. It suggests a qualitatively different behavior of the two-particle
correlation as a function of pseudorapidity sum in p+p and Au+Au collisions.

\end{abstract}

\vskip0.6cm

\textbf{1.} Correlations between particles produced in different rapidity
regions have been intensively studied since the early times of high-energy
physics \cite{ki-wo}. Particularly interesting are correlations between
particles with large separation in rapidity. It is recognized that such
correlations are born immediately after the collision, when the produced
system is very small (spatial size of the order of a few femtometers) and
before rapid longitudinal expansion.

One popular method to study long-range correlations is to measure the
multiplicity correlation coefficient, i.e., to quantify how multiplicity
(number of particles) in one rapidity window influences multiplicity in
another one. This problem was thoroughly studied in hadron-hadron collisions
at various energies \cite{isr-fb,ua5-fb,e735-fb}, \cite%
{dpm1,dpm2,bbp,fial,chou,zajc,bpv,giov,bzd-pp,bz}. One important lesson from
these studies is that the forward-backward correlation coefficient decreases
as a function of rapidity distance between bins.

Recently the STAR Collaboration at RHIC announced the results \cite{star-prl}
of the forward-backward multiplicity correlation coefficient measured in
Au+Au collisions at $\sqrt{s}=200$ GeV. The measurement was performed for
two narrow pseudorapidity bins with the distances between them ranging from $%
0.2$ to $1.8$, covering a substantial part of the midrapidity region. For
the first time very interesting features were observed: (i) the correlation
coefficient increases significantly with centrality of the collision, and
(ii) it remains approximately constant (except for very peripheral
collisions) across the measured midrapidity region $\left| \eta \right| <1$.
These results were interpreted in the framework of the color glass
condensate \cite{cgc-fb} or the dual parton \cite{dpm2} models.

Recently various mechanisms have been proposed to understand the data
quantitatively \cite{gio,bzd,paj,yan}. However, in these calculations the
sophistication of the STAR analysis was not fully appreciated, and the
published results cannot be directly compared with data. As emphasized by
Lappi and McLerran \cite{lap-mcl} in the STAR analysis, the correlation
coefficient is measured at a given number of particles in an additional
reference window. This procedure significantly influences the
forward-backward correlations, and we come back to this problem later.

In the present paper we analyze the STAR data and extend the discussion
initiated in Ref. \cite{lap-mcl}. We describe the STAR analysis in detail
and derive a general formula that relates the correlation coefficients
measured with and without the step of fixing particle number in the
reference window.

The main result of this study is the observation that the two-particle
pseudorapidity correlation function is qualitatively different in p+p and
central Au+Au collisions when studied as a function of pseudorapidity sum $%
\eta _{1}+\eta _{2}$. In a model independent way we show that bins located
asymmetrically around $\eta =0$ with a small separation in pseudorapidity
are significantly more weakly correlated than bins located symmetrically
with much larger separation. It is the first time this effect is observed.
In p+p collisions the opposite effect is observed, i.e., bins with smaller
separation are more strongly correlated even if they are asymmetric.

\bigskip

\textbf{2.} The multiplicity correlation coefficient for two bins $X$ and $Y$
is 
\begin{equation}
b_{XY}=\frac{D_{XY}^{2}}{D_{XX}D_{YY}},  \label{b_def_1}
\end{equation}%
\begin{equation}
D_{XY}^{2}=\left\langle n_{X}n_{Y}\right\rangle -\left\langle
n_{X}\right\rangle \left\langle n_{Y}\right\rangle ;\quad
D_{YY}^{2}=\left\langle n_{Y}^{2}\right\rangle -\left\langle
n_{Y}\right\rangle ^{2},  \label{b_def_2}
\end{equation}%
where $n_{X}$ and $n_{Y}$, respectively, are event-by-event multiplicities
in $X$ and $Y$. Due to the Cauchy-Schwarz inequality $b_{XY}$ varies from $%
-1 $ to $+1$.

The STAR Collaboration measured the multiplicity correlation coefficient
between two symmetric (with respect to $\eta =0$ in the center-of-mass
frame) pseudorapidity bins $B$ (backward) and $F$ (forward) of width $0.2$.
To reduce a trivial source of correlations coming from the impact parameter
fluctuations,\footnote{%
Higher $n_{B}$ triggers a smaller impact parameter that leads to higher $%
n_{F}$.} STAR introduced the third symmetric reference bin $R$ (see Fig. \ref%
{fig1}), and all averages $\left\langle n_{B}\right\rangle _{n_{R}}$, $%
\left\langle n_{B}^{2}\right\rangle _{n_{R}}$, and $\left\langle
n_{B}n_{F}\right\rangle _{n_{R}}$ were measured at a given number of
particles $n_{R}$ in this bin. Next they calculated the appropriate
covariance and variance in the following way: 
\begin{eqnarray}
D_{BF}^{2}|_{STAR} &=&\sum\nolimits_{n_{R}}P(n_{R})\left[ \left\langle
n_{B}n_{F}\right\rangle _{n_{R}}-\left\langle n_{B}\right\rangle _{n_{R}}^{2}%
\right] ,  \notag \\
D_{BB}^{2}|_{STAR} &=&\sum\nolimits_{n_{R}}P(n_{R})\left[ \left\langle
n_{B}^{2}\right\rangle _{n_{R}}-\left\langle n_{B}\right\rangle _{n_{R}}^{2}%
\right] ,  \label{D2star_1}
\end{eqnarray}%
where $P(n_{R})$ is the multiplicity distribution in the reference bin $R$
at a given centrality class that is defined by a range of $n_{R}$, i.e., $%
n_{1}<n_{R}<n_{2}$. Equation (\ref{D2star_1}) allows us to calculate the
correlation coefficient as measured by STAR: 
\begin{equation}
b_{BF}|_{STAR}=\frac{D_{BF}^{2}|_{STAR}}{D_{BB}^{2}|_{STAR}}.
\end{equation}

It is important to emphasize that if $\left\langle n_{B}\right\rangle $, $%
\left\langle n_{B}^{2}\right\rangle $, and $\left\langle
n_{B}n_{F}\right\rangle $ are measured without the step of fixing $n_{R}$
(namely all events are taken to directly measure $D_{BF}^{2}$ and $%
D_{BB}^{2} $ with $n_{R}$ in a given centrality range) different results are
obtained.\footnote{%
Naively, it seems that both procedures should lead to the same result. We
can always measure $\left\langle O\right\rangle _{n_{R}}$ at a given $n_{R}$
and calculate $\left\langle O\right\rangle
=\sum\nolimits_{n_{R}}P(n_{R})\left\langle O\right\rangle _{n_{R}}$. In this
case,%
\begin{equation*}
D_{BF}^{2}=\left\langle n_{B}n_{F}\right\rangle -\left\langle
n_{B}\right\rangle ^{2}=\sum\nolimits_{n_{R}}P(n_{R})\left\langle
n_{B}n_{F}\right\rangle _{n_{R}}-\left(
\sum\nolimits_{n_{R}}P(n_{R})\left\langle n_{B}\right\rangle _{n_{R}}\right)
^{2},
\end{equation*}%
which is clearly different from Eq. (\ref{D2star_1}).} In the following all
observables without a label \textit{STAR} denote that $D_{BF}^{2}$ and $%
D_{BB}^{2}$ are calculated without fixing $n_{R}$. 
\begin{figure}[h]
\begin{center}
\includegraphics[scale=1.1]{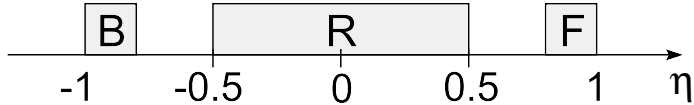}
\end{center}
\caption{Configuration with maximum pseudorapidity gap between $B$ and $F$.}
\label{fig1}
\end{figure}

The STAR procedure of measuring $b_{BF}|_{STAR}$ substantially removes the
impact parameter fluctuations, indeed. However, as shown in Ref. \cite%
{lap-mcl}, it complicates the interpretation of $b_{BF}|_{STAR}$ since it
clearly depends (in the nontrivial way) on correlations between $B(F)$ and $%
R $. In the following we derive the relation between $b_{BF}|_{STAR}$ and
multiplicity correlations $b_{BF}$ and $b_{BR}=b_{FR}$ that are obtained in
the same centrality class but without the step of fixing $n_{R}$. Such
calculation was performed in Ref. \cite{lap-mcl}, where for simplicity the
multiplicity distribution $P(n_{B},n_{F},n_{R})$ was assumed to be in a
Gaussian form. Here we show that the result derived in Ref. \cite{lap-mcl}
is independent on $P(n_{B},n_{F},n_{R})$ provided the average number of
particles in $B$ at a given $n_{R}$ is a linear function of $n_{R}$:%
\begin{equation}
\left\langle n_{B}\right\rangle _{n_{R}}=c_{0}+c_{1}n_{R}.  \label{nb_lin}
\end{equation}%
This relation is well confirmed by STAR \cite{tarn}. It is straightforward
to show that 
\begin{equation}
c_{0}=\left\langle n_{B}\right\rangle -\left\langle n_{R}\right\rangle \frac{%
D_{BR}^{2}}{D_{RR}^{2}},\quad c_{1}=\frac{D_{BR}^{2}}{D_{RR}^{2}}.
\label{c0c1}
\end{equation}%
Indeed, to obtain Eq. (\ref{c0c1}) both sides of Eq. (\ref{nb_lin}) should
be multiplied first by $P(n_{R})$ and second by $P(n_{R})n_{R}$ and summed
over $n_{R}$. Using an obvious relation 
\begin{equation}
\left\langle O\right\rangle _{n_{R}}=\frac{1}{P(n_{R})}%
\sum_{n_{B},n_{F}}P(n_{B},n_{F},n_{R})O,  \label{Odef}
\end{equation}%
two simple equations can be derived that allow us to calculate $c_{0}$ and $%
c_{1}$.

Taking Eqs. (\ref{D2star_1}), (\ref{nb_lin}), and (\ref{Odef}) into account,%
\begin{eqnarray}
D_{BF}^{2}|_{STAR} &=&D_{BF}^{2}-c_{1}^{2}D_{RR}^{2},  \notag \\
D_{BB}^{2}|_{STAR} &=&D_{BB}^{2}-c_{1}^{2}D_{RR}^{2},  \label{D2star_2}
\end{eqnarray}%
where $c_{1}$ is defined in (\ref{c0c1}). Consequently, $b_{BF}|_{STAR}$ is
given by%
\begin{equation}
b_{BF}|_{STAR}=\frac{b_{BF}-b_{BR}^{2}}{1-b_{BR}^{2}},  \label{b_STAR}
\end{equation}%
where $b_{BF}$ and $b_{BR}$ are the appropriate correlation coefficients
measured without fixing $n_{R}$. As mentioned earlier we obtain exactly the
same formula as in Ref. \cite{lap-mcl}. It shows that Eq. (\ref{b_STAR})
does not depend on $P(n_{B},n_{F},n_{R})$, provided the relation (\ref%
{nb_lin}) is satisfied.

Here point out that the interpretation of $b_{BF}|_{STAR}$ is not
straightforward. For example, $b_{BF}|_{STAR}=0$ indicates only that $%
b_{BF}=b_{BR}^{2}$ but it does not mean that $b_{BF}=0$. Moreover, $%
b_{BF}|_{STAR}$ can be negative even if both $b_{BF}$ and $b_{BR}$ are
positive. We conclude that the full interpretation of $b_{BF}|_{STAR}$ is
difficult without knowing $b_{BF}$ and $b_{BR}$.

In this paper we are interested in the configuration presented in Fig. \ref%
{fig1}, where the distance between $B$ and $F$ is a maximum one, i.e., $%
F=[0.8<\eta <1]$, $B$ is symmetric with respect to $\eta =0$, and $%
R=[-0.5<\eta <0.5]$. In this case the average gap between $B$ and $R$ is
smaller by a factor of $2$ than that between $B$ and $F$. Assuming that the
two-particle correlation function depends only on $\left| \eta _{1}-\eta
_{2}\right| $ and is not increasing as a function of $\left| \eta _{1}-\eta
_{2}\right| $ a \textit{natural} ordering $b_{BR}\geq b_{BF}$ is obtained,
as shown explicitly in Ref. \cite{lap-mcl}. Consequently%
\begin{equation}
b_{BF}|_{STAR}=\frac{b_{BF}-b_{BR}^{2}}{1-b_{BR}^{2}}\leq \frac{%
b_{BR}-b_{BR}^{2}}{1-b_{BR}^{2}}=\frac{b_{BR}}{1+b_{BR}}\leq \frac{1}{2},
\label{limit}
\end{equation}%
since $b_{BR}\leq 1$. In the most central collisions STAR measured $%
b_{BF}|_{STAR}\approx 0.58$, which violates this bound.\footnote{%
The STAR result has an uncertainty $\pm 0.06$. Even if one assumes that the
measured $b_{BF}|_{STAR}$ is slightly below $0.5$, it is still difficult to
understand with an assumption $b_{BR}\geq b_{BF}$, since it requires $%
b_{BR}\approx b_{BF}\approx 1$.} Thus we arrive at an interesting conclusion
that in the midrapidity region in the most central Au+Au collisions the
following inequality holds:%
\begin{equation}
b_{BR}<b_{BF}.  \label{ineq}
\end{equation}

It was checked by STAR that narrowing the reference bin $R$ from $\left|
\eta \right| <0.5$ to $\left| \eta \right| <0.1$ (so that all windows have
the same widths) slightly increases the correlation coefficient $%
b_{BF}|_{STAR}$. Also an alternative method of centrality determination was
carried out using the STAR zero-degree calorimeter (measurement of forward
neutrons) for the $0-10\%$ centrality, and $b_{BF}|_{STAR}$ is very close to 
$\frac{1}{2}$. In this case the same formula (\ref{D2star_1}) applies;
however, there are no explicate cuts on $n_{R}$. We conclude that the width
of $R$ and the centrality cut on $n_{R}$ is not a factor in the result (\ref%
{ineq}).

\bigskip

\textbf{3.} It is interesting to estimate the numerical values of the
correlation coefficients $b_{BF}$ and $b_{BR}$. As mentioned earlier we are
mostly interested in the configuration where the distance between $B$ and $F$
is a maximum one ($\Delta \eta =1.8$ in the STAR notation) and $R$ is
defined by $\left| \eta \right| <0.5$.

As seen from Eq. (\ref{D2star_2}) evaluation of $%
b_{BF}=D_{BF}^{2}/D_{BB}^{2} $ is straightforward. The covariance $%
D_{BF}^{2}|_{STAR}$ and variance $D_{BB}^{2}|_{STAR}$ are published in \cite%
{star-prl} (only for $0-10\%$ centrality bin). From Ref. \cite{tarn} one
sees that $\left\langle n_{B}\right\rangle _{n_{R}}$ is a linear function of 
$n_{R}$ with a coefficient $c_{1}\approx 0.2$. To calculate $%
D_{RR}^{2}=\left\langle n_{R}^{2}\right\rangle -\left\langle
n_{R}\right\rangle ^{2}$ we use the uncorrected (raw) multiplicity
distribution $P(n_{R}^{\text{raw}})$ as published in Ref. \cite{star-all},
and take the efficiency correction to be $n_{R}/n_{R}^{\text{raw}}=1.22$ %
\cite{tarn,star-all}. Performing a straightforward calculation we obtain%
\footnote{%
We take $P(n_{R}^{\text{raw}})\propto \exp (-\frac{n_{R}^{\text{raw}}}{370})$
for $431\leq n_{R}^{\text{raw}}\leq 560$ and $P(n_{R}^{\text{raw}})\propto
\exp (-\frac{(n_{R}^{\text{raw}}-561)^{2}}{2700})$ for $n_{R}^{\text{raw}%
}\geq 561$, which gives $D_{RR}^{2}|_{\text{raw}}=2904$. Consequently, $%
D_{RR}^{2}=(1.22^{2})D_{RR}^{2}|_{\text{raw}}$.} $D_{RR}^{2}\approx 4320$,
which allows us to calculate $b_{BF}$. Taking Eq. (\ref{b_STAR}), $b_{BF}$,
and measured $b_{BF}|_{STAR}$ into account we obtain%
\begin{equation}
b_{BR}\approx 0.58,\quad b_{BF}\approx 0.72.  \label{nume}
\end{equation}%
As seen from (\ref{nume}) in the most central Au+Au collisions $b_{BR}$ is
significantly smaller than $b_{BF}$. Let us note here that the average
distance between $B$ and $R$ (one unit of $\eta $) is smaller by a factor of
two than that between $B$ and $F$.

It is also interesting to see how $b_{BF}$ depends on the distance $\Delta
\eta $ between bins $B$ and $F$. Taking Eq. (\ref{D2star_2}) into account
and repeating calculations\footnote{%
For small $\Delta \eta $ the reference window $R$ is composed of two windows 
$0.5<\left| \eta \right| <1$ and we assume that $c_{1}^{2}D_{RR}^{2}$ is
approximately the same as with $R$ defined by $\left| \eta \right| <0.5$.}
presented above we found that $b_{BF}$ in central Au+Au collisions is
approximately constant as a function of $\Delta \eta $, which is consistent
with the dependence of $b_{BF}|_{STAR}$ on $\Delta \eta $.

Finally, let us notice that STAR also measured $b_{BF}|_{STAR}$ in p+p
collisions; however, in this case the exact value of $c_{1}$ is not known.
We checked that for a very broad range of $c_{1}$ we always obtain a
standard ordering $b_{BR}>b_{BF}$.\footnote{%
We assume $P(n_{R})$ to be given by a negative binomial distribution with
standard parameters $\left\langle n_{R}\right\rangle =2.3$ and $k=2$.
Taking, e.g., $c_{1}=0.1$ we obtain $b_{BR}\approx 0.28$ and $b_{BF}\approx
0.13$.}

\bigskip

\textbf{4.} Several comments are warranted:

(i) To calculate the correlation coefficients $b_{BF}$ and $b_{BR}$ the
experimental values of $D_{BF}^{2}|_{STAR}$ and $D_{BB}^{2}|_{STAR}$ are
required as an input. Unfortunately they are provided only for the most
central collisions. It would be interesting to measure the centrality
dependence of the effect reported in this paper. It is expected that in
peripheral collisions the standard relation $b_{BR}>b_{BF}$ should be
recovered. If so, it would indicate a qualitatively different behavior of
central and peripheral Au+Au collisions.

(ii) It is worth mentioning that HIJING \cite{Wang:1991hta} and the Parton
String Model (PSM) \cite{Amelin:2001sk} fail to describe the Au+Au data for
the forward-backward multiplicity correlation coefficient. However, they are
consistent with the p+p data. In the most central Au+Au collisions, and for
the configuration presented in Fig. \ref{fig1}, both models predict $%
b_{BF}|_{STAR}<\frac{1}{2}$, which is consistent with the relation $%
b_{BR}>b_{BF}$.\footnote{%
In particular $b_{BF}|_{STAR}\approx 0.1$ in HIJING and $b_{BF}|_{STAR}%
\approx 0.4$ in PSM, see Ref. \cite{star-prl}.}

(iii) It is not straightforward to propose a realistic mechanism that more
strongly correlates bins $B$ and $F$ than bins $B$ and $R$. One possible
mechanism is the formation of certain \textit{clusters} strongly peaked at $%
\eta =0$ that decay symmetrically into two particles. This mechanism
obviously correlates bins $B$ and $F$ and introduces no (or much weaker)
correlations between bins $B$ and $R$. To go beyond speculations more
detailed measurement of the forward-backward correlations between symmetric
and asymmetric bins is warranted.

\bigskip

\textbf{5.} In summary, we analyzed the STAR data on the forward-backward
multiplicity correlation coefficient $b_{BF}|_{STAR}$ in the most central
Au+Au collisions. This measurement was performed with the intermediate step
of fixing the number of particles in the third reference window $R$, see
Fig. \ref{fig1}, and we emphasized the importance of this step. We derived
the general formula that relates $b_{BF}|_{STAR}$ and the correlation
coefficients $b_{BF}$ and $b_{BR}$ measured in $B-F$ and $B-R$ without
fixing the number of particles in $R$.

The most important result is the observation that for the configuration
presented in Fig. \ref{fig1}; in the most central Au+Au collisions, the
correlation coefficient $b_{BR}$ is significantly smaller than $b_{BF}$.
This is exactly opposite of what is expected and measured in p+p collisions
(the distance between $B$ and $R$ is smaller by a factor of $2$ than that
between $B$ and $F$). Moreover, we found that in central Au+Au collisions, $%
b_{BF}$ is approximately constant as a function of the pseudorapidity
separation between symmetrically located bins $B$ and $F$. To understand
these results it is necessary to assume that in central Au+Au collisions the
two-particle correlation function strongly decreases as a function of $|\eta
_{1}+\eta _{2}|$. It indicates the presence of a specific mechanism of
correlation that strongly correlates bins located symmetrically around $\eta
=0$ for which $|\eta _{1}+\eta _{2}|\approx 0$, but is less effective for
asymmetric bins $|\eta _{1}+\eta _{2}|>0$.\footnote{%
It also indicates a strong violation of boost invariance in the midrapidity
region \cite{Bialas:2011bz}.}

In this paper we solely concentrated on an analysis of the experimental
results and at the moment we see no compelling explanation of this effect.
It would be interesting to directly measure at RHIC and LHC the multiplicity
correlation coefficient for symmetric and asymmetric bins to confirm
conclusions presented in this paper.

\bigskip

\textbf{Acknowledgments}

We thank Andrzej Bialas and Larry McLerran for enlightening discussions.
Correspondence with Brijesh Srivastava is highly appreciated. This
investigation was supported by the U.S. Department of Energy under Contract
No. DE-AC02-98CH10886 and by Grant No. N202 125437 of the Polish Ministry of
Science and Higher Education (2009-2012).

\end{document}